\documentclass[10pt]{iopart}

\usepackage[utf8]{inputenc} 
\usepackage[T1]{fontenc}    
\usepackage{hyperref}       
\usepackage{url}            
\usepackage{booktabs}       
\usepackage{amsfonts}       
\usepackage{nicefrac}       
\usepackage{microtype}      
\usepackage{natbib}
\usepackage{graphicx}

\newcommand{\apjs}{ApJS}
\newcommand{\apjl}{ApJL}

\newcommand{\aap}{A\&A}

\begin{document}

\title[HiFLEx]{HiFLEx -- a highly flexible package to reduce cross-dispersed echelle spectra}

\author{
  Ronny Errmann$^1$, 
  Neil Cook$^2$,
  Guillem Anglada-Escudé$^3$,
  Sirinrat Sithajan$^4$, 
  David Mkrtichian$^4$,
  Eugene Semenko$^4$,
  William Martin$^1$,
  Tabassum S. Tanvir$^1$,
  Fabo Feng$^1$,
  J.L. Collett$^1$ and
  Hugh R. A. Jones$^1$
  }
\address{$^1$The Centre for Astrophysics Research, Astronomy and Mathematics School, University of Hertfordshire, Hatfield, Hertfordshire, AL10 9AB, UK\\
$^2$ Institut de Recherche sur les Exoplanètes, Département de Physique, Université de Montréal, Montréal, QC H3C 3J7, Canada\\
$^3$ Institut de Ciències de l'Espai Barcelona, 08035, Spain\\
$^4$ National Astronomical Research Institute of Thailand (NARIT), 260 Moo 4, T. Donkaew, A. Maerim, Chiangmai, 50180, Thailand\\}


\begin{abstract}
We describe a flexible data reduction package for high resolution cross-dispersed echelle data. This open-source package is developed in \texttt{Python} and includes optional GUIs for most of the steps. It does not require any pre-knowledge about the form or position of the echelle-orders. It has been tested on cross-dispersed echelle spectrographs between 13k and 115k resolution (bifurcated fiber-fed spectrogaph ESO-HARPS and single fiber-fed spectrograph TNT-MRES). HiFLEx can be used to determine radial velocities and is designed to use the TERRA package but can also control the radial velocity packages such as CERES and SERVAL to perform the radial velocity analysis. Tests on HARPS data indicates radial velocities results within $\pm3$\,m/s of the literature pipelines without any fine tuning of extraction parameters.

\end{abstract}

\noindent{\it Radial velocity–Spectroscopy–Astronomy–data reduction}

\noindent{Published in Publications of the Astronomical Society of the Pacific (PASP), 132:064504(10pp), 2020 June }


\section{Introduction}
High resolution echelle spectrographs are a highly important tool to study astrophysical phenomena like the study of stellar atmospheres (e.g. \cite{2019A&A...631L...3R}) or the search for and characterisation of extrasolar planets around nearby stars (e.g. \cite{2016Natur.536..437A, 2017A&A...599A..16L}). 
High resolution is reached by using an echelle grating, which is optimised for reflection in high orders, as higher orders have higher dispersion. However, these orders are overlapping in space, therefore a cross disperser is used to separate the orders with dispersion perpendicular to the echelle dispersion.

For precise results, instruments require precise data reduction packages. Many of the current high resolution spectrographs have dedicated data reduction packages, e.g. DRS (Data Reduction System) for HARPS and ESPRESSO \citep{2018SPIE10704E..0FD, 2015ASPC..495..285S}, CARACAL for CARMENES \citep{2016SPIE.9910E..0EC}, or APERO for SPIRou/NIRPS \citep{Cook20}. However, for some echelle spectrographs no data reduction pipeline is available.

Additional to the data reduction packages provided for a single instrument, more general data reduction packages which can be applied to a larger set of echelle instruments are available. Packages to extract data from stabilised spectrographs are available (e.g. \citet{2014A&A...561A..59Z}), however, only some echelle spectrographs are stabilised. Another example is the CERES pipeline \citep{2017PASP..129c4002B}, which allows the reduction of data from HARPS \citep{2003Msngr.114...20M}, FEROS \citep{1999Msngr..95....8K}, HIRES \citep{1994SPIE.2198..362V}, CORALIE, and several other spectrographs. However, for each instrument the pipeline requires a bespoke script with several hundred lines of code and prepared configuration files with the positions of the orders and the wavelength solution. This makes for quite a lot of extra steps to apply it to data from spectrographs not included. The preset requirements make data reduction for an instrument in development, when the position of the echelle orders or resolution might change often, a very considerable overhead.

We are developing a new high resolution spectrograph \citep{10.1117/12.2529044} for the Thai National Telescope (TNT). The spectrograph will be fiber-fed with a bifurcated fiber input and a foreseen resolution of 100,000. We test a variety of different designs. 
The available pipelines were not sufficiently flexible to enable quick analysis of our data as many variations in the code were required for each new setup. Instead we developed a flexible data reduction package, in which a single calibration file has to be edited while no modification in the program code are required to reduce data from a modified setup. This flexible package also allows the reduction from data of a wide range of echelle spectrographs. We tested the package on data from established spectrographs.

After the extraction and calibration of the spectrum, precise radial velocities can be measured using available packages: TERRA \citep{2012ApJS..200...15A}, SERVAL\footnote{\url{https://github.com/mzechmeister/serval}} \citep{2018A&A...609A..12Z}, and CERES pipeline\footnote{\url{https://github.com/rabrahm/ceres}}. The description of these packages can be found in the corresponding publications. Further radial velocity packages are available to analyse the reduced spectra, e.g. \citet{2019AJ....158..164B, 2020MNRAS.492.3960R}. Increase precision of the barycentric corrections can be done with packages like PEXO \citep{Feng_2019}.

\section{Description of the package}
\label{sec:describtion_package}
\begin{figure}
  \centering
  \includegraphics[width=0.5\textwidth]{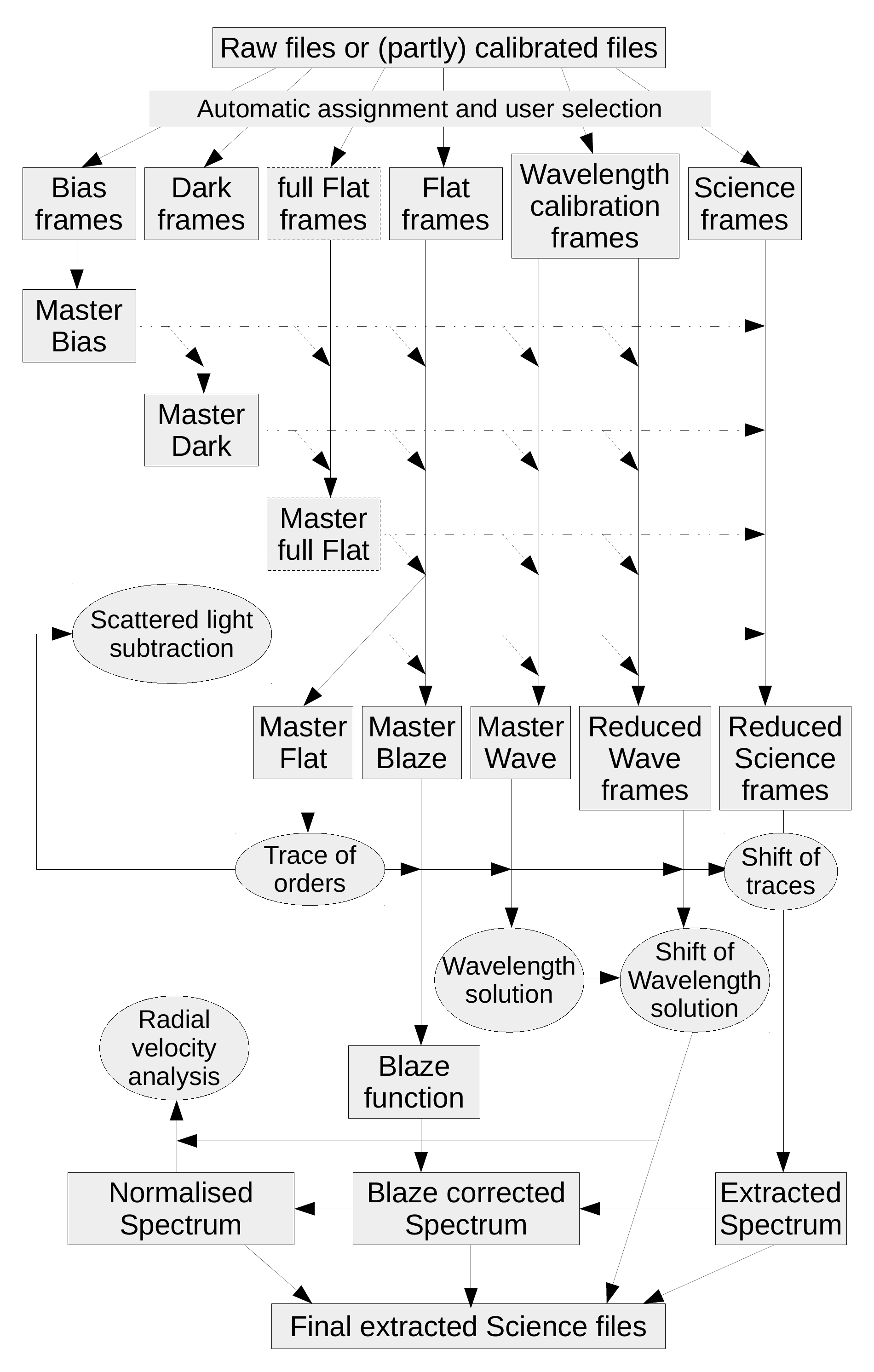}
  \caption{Flow chart of the data reduction and analysis of HiFLEx. The chart shows all available steps for data processing and a suggested usage of calibration data. Wave stands for the images from which a wavelength solution can be determined, e.g. an emission lamp spectrum. However, the pipeline also works with subsets of these steps. For example, taking full frame flats (the detector is illuminated uniformly) is not possible for most spectrographs, hence this step is most often skipped. On the other hand, if only a quick glance of an uncalibrated (no wavelength solution) spectrum is required, then a single science frame is enough, which can be used as master flat and master blaze. Everything between this extreme and full calibration is possible.}
  \label{fig:data_reduction_flow}
\end{figure}

The purpose of the HiFLEx pipeline is to reduce and analyse echelle spectra with minimal human interaction. The pipeline works for echelle spectrographs with either single fiber or bifurcated fiber input if at least two distinguishable echelle orders are imaged onto the detector. 
The data reduction package is written in \texttt{Python} and has a modular setup. The \texttt{numpy} and \texttt{scipy} packages of \texttt{Python} are implemented for most calculations. The used fitting routines are based on the least square minimization Levenberg-Marquardt algorithm (for polynomial or Gaussian fits) or the \texttt{dgelsd} routine  which uses Singular value decomposition (for two-dimensional polynomial fits).  HiFLEx can be downloaded from Github\footnote{\label{Footnote:url_github_exohspec}\url{https://github.com/ronnyerrmann/HiFLEx/}}. In this section we describe the essential steps carried out by the pipeline, for the practical application we refer to the manual\footnote{\label{Footnote:url_github_manual}\url{https://github.com/ronnyerrmann/HiFLEx/blob/master/HiFLEx_UserManual.pdf}}. The flow chart of the package is described in Figure~\ref{fig:data_reduction_flow} and will be described in this chapter.

\subsection{Preparation}
All configuration can be done in a configuration file. A description of the available parameters can be found in the manual provided with the package and in the configuration file itself. In Table~\ref{Tab:default_parameters_conf} we provide a list with the most important parameters to be modified to set up the extraction of a new spectrograph.

\begin{table}
  \caption{List of the most important parameters from the configuration file to set up HiFLEx for a new spectrograph. Some suggested default values are given in italic. Detailed information can be found in the manual provided with the pipeline$^{\ref{Footnote:url_github_exohspec}}$. In order to run the pipeline, 'run \texttt{file\_assignment.py}' (to check and modify assignments of raw files to calibration files or reduction steps) and then 'run \texttt{hiflex.py}' (to perform the calibration steps and extract the science spectra).}
  \label{Tab:default_parameters_conf}
  \begin{tabular}{l l }
   Parameter        & Short description   \\
   \hline\hline
   GUI              & \textit{True} [modify once appropriate setup has been found] \\ \hline
   \texttt{raw\_data\_paths} & Folders of the raw data \\ \hline
   \texttt{rotate\_frame}, & To orientate the raw images: vertical traces, \\
   \texttt{flip\_frame}    & reddest pixel at (0,0)  \\ \hline
   \texttt{arcshift\_side} & Calibration traces are \textit{left} to, \textit{right} to, or at the \\
                           & same (\textit{center}) position as science traces \\ \hline
   \texttt{original\_master\_wave}- & Set to new file to use a GUI to create new \\
   \texttt{lensolution\_filename}   & wavelength solution (or set to previous solution) \\ \hline
   \texttt{raw\_data\_}*            & Set the keyword names and values to assign files \\
                                    & to reduction steps automatically \\ \hline
   *\texttt{\_calibs\_create\_g}    & Define the processing of the raw files for \\
                                 & each image type or reduction step \\ \hline
   \texttt{altitude}, \texttt{latitude}, \texttt{longitude}  & Observation site, if not in header \\ \hline
   \texttt{terra\_jar\_file}, \texttt{path\_ceres}, & Paths to external radial velocity packages \\
   \texttt{path\_serval}                   & (optional, need to be downloaded manually)\\ \hline
   \end{tabular}
\end{table}

The package assigns the available frames to the calibration steps depending on the header of the files and the name of the file using a common naming convention, which is given in the manual. Afterwards the pipeline allows the user to modify the assignment of the individual observation images for the different calibration steps.

\subsection{Image processing and master files}
The calibration of the image frames depends on the instrument. For example, a CCD detector cooled to liquid nitrogen temperatures might not require dark frame correction when observing bright targets, while dark correction is highly recommended for a Peltier cooled frame from a CMOS architecture detector.

The HiFLEx package allows the user to apply a broad variety of calibration steps. Additionally, pre-reduced images or master images can be given, to which none or only some calibration steps will be applied. Possible calibrations include windowing, bad-pixel mask correction for dead pixels, correction with dark, bias, or flat field images, and removal of background (scattered) light measured between the orders. 

If the user provides a bad-pixel mask, these bad pixels will be replaced by the median value of the surrounding pixel, using a 3x3 box around the bad pixel. Extracted spectral regions that contained bad pixels during the extraction will be marked so they can be excluded from further analysis.


The master files (e.g. Bias, Dark, Flat-field, or image to trace the orders or extract the blaze function) are created by the pipeline once they are required using the (raw-) frames and calibration steps as given by the user. The images are median-combined on a pixel-by-pixel basis. However, the user can also combine images using the sum or average. Before any combination, images can also be scaled to the average median flux of the frames before combining, e.g. if the light source varies in time. Master frames already created by the pipeline are automatically used or those specified by the user will be read.

\subsection{Finding and defining the traces of the echelle orders}

\begin{figure}
  \centering
  \includegraphics[width=0.5\textwidth]{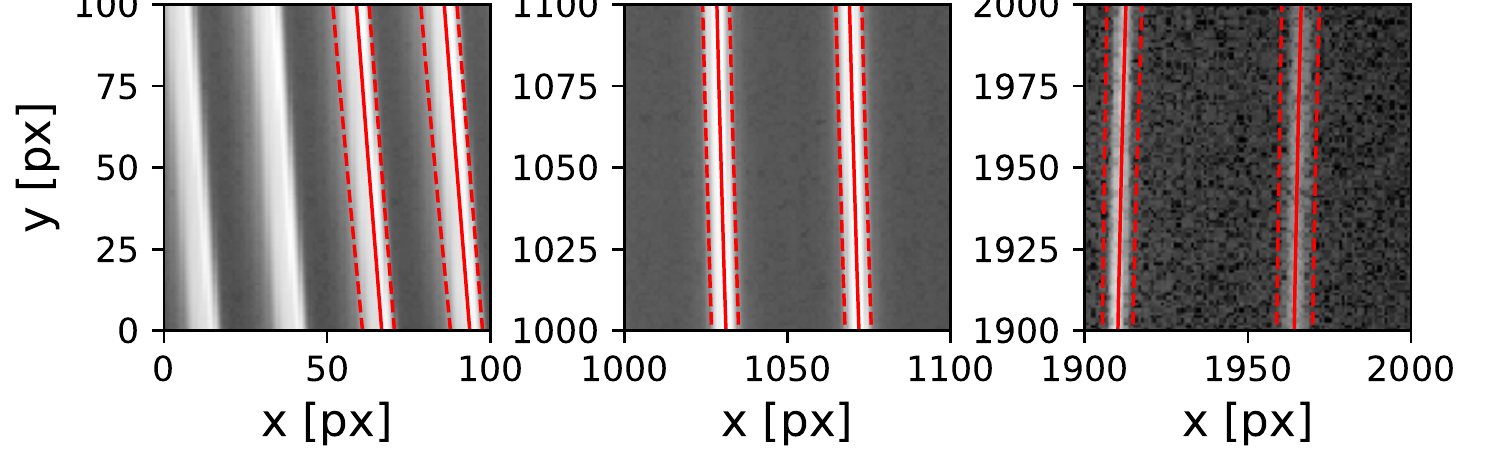}
  \caption{HiFLEx identified orders in three different areas of the detector: two opposite corners (left and right panel) and the centre (middle panel).
  The data was taken with the echelle spectrograph in development for the TNT. The centre of the order is marked with a solid line, the extraction area (the flux is above 5\% of the maximum flux) with dashed lines. Two orders in the left panel are not identified, as only a small portion of the orders are presented in the image. The whole arrangement of orders is shown in Figure~\ref{fig:ThAr_traces_image} with the same pixel labels. }
  \label{fig:traces_image}
\end{figure}

While it is a relatively easy task for humans to define the traces of echelle spectra, this task is harder to program in a general way. For example in some spectrographs the red orders are partly overlapping or the blue orders have a much smaller signal than the stray light between the red orders. To keep flexibility in our package, the ability to trace the orders should not depend on the curvature of the orders.
For this step the user can use any data, however, we suggest to use a master file created from several high signal-to-noise continuum spectra (e.g. Tungsten) in the science fiber. In the following explanation we assume that the traces of the orders are vertically oriented.

Our approach for finding the traces is two-step. At first, in a binned image the traces are searched. Binning increases the signal-to-noise ratio and decreases the calculation time. In tests we saw that using the mean instead of the median for binning allows a better tracing, especially when large bin sizes in the cross-dispersion direction can be used due to large gaps between the traces of the orders. 
Second, the iterative finding of all orders is done. The pixel with the highest flux is searched and a Gaussian function fitted to the 1-dimensional cross section in the cross-dispersion (horizontal) direction (with a width 1 pixel in the vertical direction). The centre of the Gaussian fit is assumed to be the centre of the trace at this dispersion pixel. The Gaussian fit is then repeated on 1-dimensional cross section above and below the previous position, using the previous fitted centre as the starting point until the fit fails on several consecutive cross sections, e.g. because the Gaussian fit is less significant than $3\,\sigma$. If the information from other traces is available, then this information is used to follow the curvature of the trace in areas where no fit is possible. Once the trace has been identified, a polynomial fit from user-defined polynomial degree $N_{\rm poly}$ (default 3) is fitted to the centre of the Gaussians $px_{cen}$ for each pixel in the trace $px_{y}$:
\begin{equation}
\label{Eq:traces_poly}
px_{\rm cen} = \sum_{i=0}^{N_{\rm poly}} a(i)\,(px_{y})^{i} \hspace{1cm},
\end{equation}
with $a(i)$ as the coefficients of the polynomial. The lowest polynomial degree to fit the data should be chosen for $N_{\rm poly}$ to avoid over-fitting. The results can be checked in the GUI or the logged images. Traces will only be used, if they were identified in more than 25\% of the image and if they do not cross previously found traces. The algorithm has been tested in images with cosmic rays and ghosts, as well as in images with high scattered light between the traces. The identification is robust against these effects. After one order has been traced, the trace of the order is masked in the binned image and the next brightest pixel searched to repeat the above steps.

The second step consists of redefining the traces and finding the extraction width of the traces. This step can be done on an image that is binned slightly in dispersion direction to speed up the process. For the previously found traces the Gaussian fit in cross-dispersion direction is repeated. The brightest area of the cross-section is also fitted with a third order polynomial to define the centre of the trace for non-Gaussian profiles. Additionally the borders of the trace, where the flux drops to a user-defined level (default 5\%) are measured by applying a linear regression to the transition area. These measurements are repeated for each cross section of the trace in dispersion (vertical) direction. The final trace and the extraction borders are defined by a polynomial fit of user-defined polynomial degree (default 3) along the centre of the Gaussian and along the measured position of the borders for each side. Figure~\ref{fig:ThAr_traces_image} shows an image with the marked traces  and Figure~\ref{fig:traces_image} a zoom showing the extraction borders (red-dashed lines). 

For the case of a bifurcated fiber feed, the traces of the calibration fiber will be determined by shifting the position of the previously identified traces in images in which the calibration fiber is illuminated. The shift between the traces of the science order and its corresponding calibration order depends on the dispersion of the cross-disperser (especially when a prism is used). For resilience against badly identified shifts which can happen when orders have very low signal, a second order polynomial is fitted to the shift of each calibration trace ($s_{o}$):
\begin{equation}
\label{Eq:traceshift_poly}
s_{o} = \sum_{i=0}^{2} a(i)\,o^{i} \hspace{1cm},
\end{equation}
with $o$ as order of the trace. Sigma-clipping is applied to remove outliers from the fit (default 3\,sigma). The final shifts are then derived from Equation~\ref{Eq:traceshift_poly}. These shifts are added to the constant term in the trace of each order, while the higher order polynomial terms are kept the same to preserve the curvature of the traces.

\subsection{Wavelength solution}
Once the position of the traces is known, a spectrum for the wavelength calibration can be extracted. The positions of these lines will then be compared with a standard list of wavelengths to find the relation between pixel-position and the wavelengths. The pipeline will use a user-defined line list for the reference wavelengths. A line list for Th-Ar and U-Ne is provided. Both lists were taken from the NIST database\footnote{\url{https://physics.nist.gov/PhysRefData/ASD/lines_form.html}} \citep{NIST}. Any other line list can be provided by the user. For example, if a Fabry-Perot interferometer or laser frequency combs is used to perform the wavelength calibration. The usage of an equally spaced wavelength source requires knowledge of the approximate wavelength calibration within a few pixel beforehand, otherwise the line list can be used as any other line list from emission lamps.

For best precision, the user should limit the line list to stable lines and the line list should be cleared of blended and asymmetric lines to avoid these lines affecting the precision of the wavelength solution. For example in the case of a spectrograph that covers the spectral range of HARPS, the HARPS catalogue can be used \citep{2007A&A...468.1115L}. 
For naming convention, we will assume from here on an emission line spectrum from a Th-Ar lamp. 

The dispersion $\theta_m$ of light of wavelength $\lambda$ is defined by the grating used and follows Eq.\,\ref{Eq:grating_equation}, with the spacing $d$ between the groves of the grating, order number $m$, and incident beam angle $\theta_i$:
\begin{equation}
\label{Eq:grating_equation}
d \left( \sin \theta_i - \sin \theta_m \right) = m\lambda \hspace{1cm}.
\end{equation}
For a single order (dispersion direction) Equation~\ref{Eq:grating_equation} leads to $\lambda \propto \sin \theta_m$, which can be approximated by a low-order polynomial to correlate pixel and wavelengths. Using a low-order polynomial also corrects distortions that might be introduced by the lens system for off-axis beams or other effects introduced by the optical system. 
In case of constant $\theta_m$ (same pixel in each order, cross-dispersion direction) the wavelength $\lambda$ in each order follows $\lambda \propto \frac{1}{m}$ (Equation~\ref{Eq:grating_equation}). However, the lens and cross-disperser might cause deviation from this equation due to their chromaticity, hence a higher-order polynomial might be necessary to correlate the order number and central wavelength. 
Finally, the correlation between inverse number of the order $\frac{1}{m}$, pixel-position $px$, and wavelength $\lambda$ will be done with a two-dimensional polynomial:
\begin{equation}
\label{Eq:poly_wave}
\lambda(px, m) = \sum_{i=0}^{N_{\rm poly,C}} \sum_{j=0}^{N_{\rm poly,D}} a(i,j) \left(\frac{1}{m}\right)^i px^j \hspace{1cm},
\end{equation}
with $a(i,j)$ as coefficients of the polynomial. The user can define the degree of the polynomial along the order-number ($N_{\rm poly,C}$) and the pixel-position ($N_{\rm poly,D}$).

The position of the emission lines in the extracted spectrum are searched in a similar way as the position of the traces. For each order a high order polynomial is fitted to the extracted flux. 
At all locations with significant remaining flux, the precise centre together with other line properties is determined by fitting a Gaussian function. A list with all the emission line positions, widths, and heights is created in this way. This list can be cleared by the user to allow removal of blended lines.

In the next step the found lines have to be compared with the set of reference lines for the corresponding lamp. If a suitable wavelength solution exists, the pipeline continues with the step described in Section~\ref{sec:adjusting_previous_wavelength_solution}. 

\subsubsection{Creation of a new wavelength solution}
The first time a completely new setup or instrument is used the user has to provide a file with order, pixel, and wavelength as an initial guess for the wavelength solution. This step can be done in a GUI that shows the list of found emission lines, as well as the visualisation of the extracted (Th-Ar) spectrum. The input will be fit with a user-defined two-dimensional polynomial (Equation~\ref{Eq:poly_wave}, default 3 for both dimensions). A suitable solution is then improved. 

\subsubsection{Adjusting from a previous wavelength solution}
\label{sec:adjusting_previous_wavelength_solution}
For most spectrographs the wavelength solution between individual nights will be nearly constant with only small changes. For stabilised spectrographs the variation will be in the sub-pixel regime. However, for the purpose of spectrograph testing the pipeline provides the user with the possibility to adjust a previous wavelength solution with a wide set of options. In case a different number of orders are identifiable between different nights, a user-defined offset range for physical dispersion order can be given. A search for a shift in dispersion direction and a variation of resolution can be performed too.

\begin{figure}
  \centering
  \includegraphics[width=0.5\textwidth]{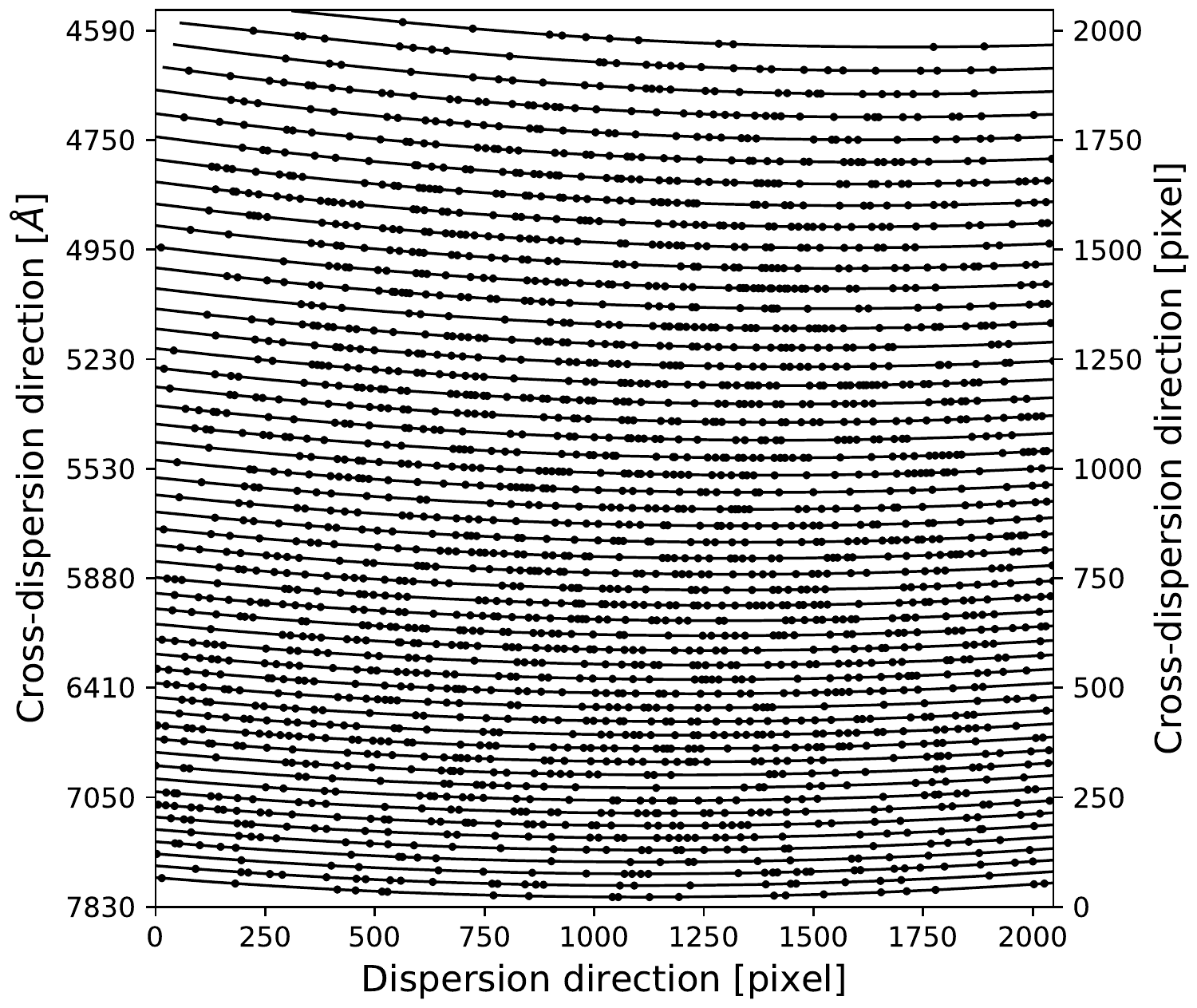}
  \includegraphics[width=0.5\textwidth]{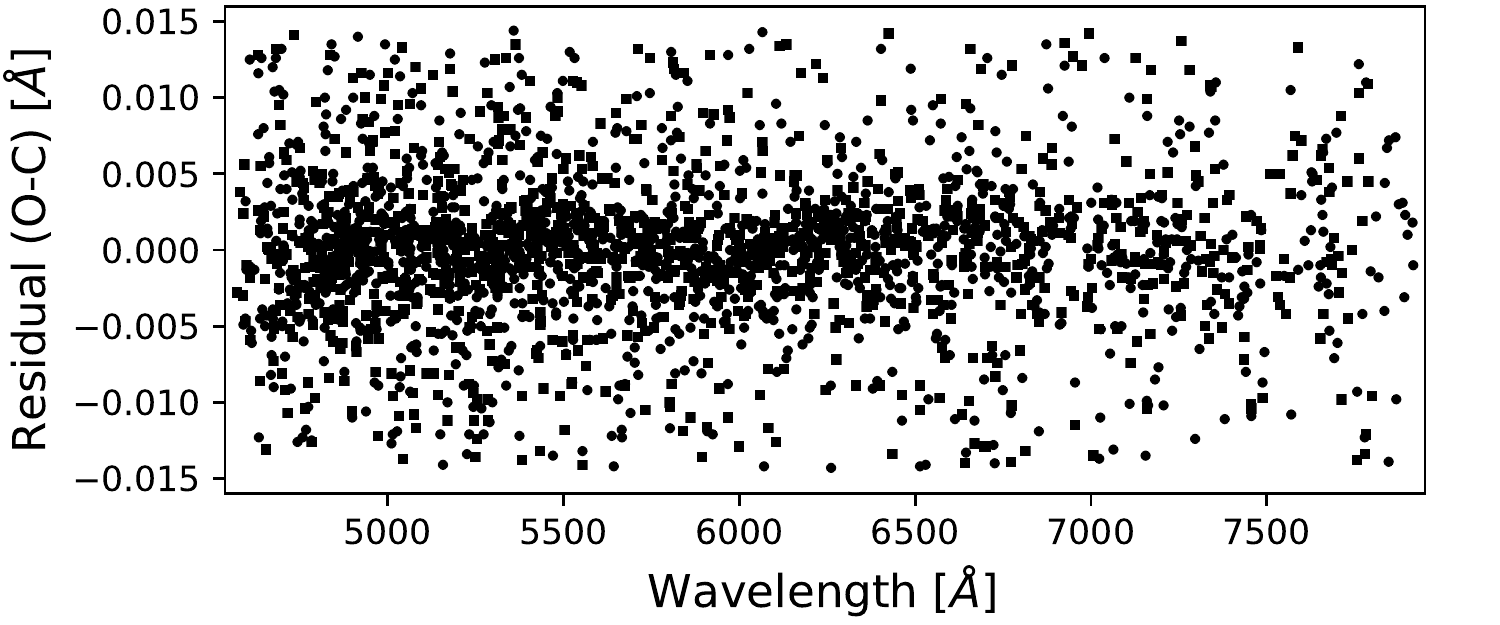}
  \caption{Top: Traces of the orders and position of the Th-Ar lines used for the wavelength solution. 
  Bottom: Residuals (Observed - Calculated) from the fit of the wavelength solution in Angstrom, shown for different wavelengths.  
  The data was taken with the prototype of an echelle spectrograph.}
  \label{fig:ThAr_traces_image}
\end{figure}

For each step in these multidimensional variations the number of correlated emission lines between the spectrum of the calibration frame and the catalogue is calculated. 
For the best matching settings, the new wavelength solution is created by fitting the correlation between inverse number of the order $\frac{1}{m}$, pixel-position $px$, and wavelength $\lambda$ with a two-dimensional polynomial (Eqation~\ref{Eq:poly_wave}) of user-defined polynomial degree (default 3 for both dimensions). The step is done iteratively, removing the least matching correlation between emission lines in the spectrum and reference catalogue. Figure~\ref{fig:ThAr_traces_image} shows an image with the marked emission lines.

\subsection{Drift of the wavelength solution}
For spectrographs that don't use simultaneous wavelength monitoring (like Th-Ar or an iodine cell) the drift of the spectrum can be measured only before and after an science exposure (or less often to avoid the loss of science time). Dealing with the drift is discussed in \citet{2014MNRAS.439.3094B}. HiFLEx provides the option to use calibration spectra to correct any drift of the spectrum. In this case the wavelength solution will be interpolated using measurements closest before and after the exposure. 

For bifurcated-fiber spectrographs the wavelength solution can be adjusted using the simultaneous Th-Ar light from the calibration fiber. Furthermore, the shift in the wavelength solutions between science and calibration fiber can be determined if images with Th-Ar in both fibers are available. The wavelength solution will be determined for both fibers and the coefficients in Equation~\ref{Eq:poly_wave}. adjusted for each solution and the shift determined from the differences. This shift is interpolated using the measurement adjacent to the science exposure and applied to the wavelength solution of the science spectrum.

\subsection{Blaze function}
\label{Section:blaze_function}
The efficiency of the echelle grating varies along each order, with the highest efficiency in the centre of the order and decreasing efficiency at the corners. This is called the blaze function. Additionally, the throughput of the whole optical system depends on the wavelength. In order to correct this effect a high signal-to-noise spectrum of a standard star or a continuum source can be used. HiFLEx allows the use of an extracted continuum spectrum (e.g. for stabilised spectrographs) or fits a polynomial along the extracted flux in each order to obtain the blaze function. Prior to fitting, the spectral features are sigma-clipped (default 4\,$\sigma$).  


\subsection{Extracting science spectra}
Before extracting the science spectra, the user-defined data reduction such as dark, bias, or background correction are performed. The background correction allows removal of stray light by measuring the signal between the orders and fitting a low order two-dimensional polynomial.
HiFLEx can also account for any drift of the spectrum in the cross-dispersion direction in the next step, where all traces are shifted to find the position in which the highest flux is extracted. To speed up the process, only the light of the central column of pixels of the shifted traces is used. To find the best shift, a Gaussian function is fitted to the total extracted flux of each shift.

The light from the science fiber is extracted from the reduced image using the shifted traces. The light for each dispersion step pixel is summed up. For the extraction of the flux at the borders of the traces the user has the option to either use the fractional flux of the pixel or the interpolated flux. The interpolated flux will produce more precise results, however it requires more computational time. Furthermore, if the border of the extraction area is at the level where the light drops down to only few percent of the maximum of this dispersion step, the improvements are typically less than 1\,ppm.

In the case of simultaneous observation of an emission line spectrum in the calibration fiber, this spectrum is extracted using the same shift as determined for the science traces. This spectrum is then used to determine the shift between the emission lines and the wavelength solution by fitting a Gaussian function to all lines used to define the wavelength solution. 
The master shift of the wavelength solution is calculated from the median of the re-identified emission lines.

In case of well exposed orders, the uncertainties of the extracted flux are dominated by the photon noise ($\sqrt{flux}$). For orders with low flux-levels additional noise sources become important. This includes the readout noise, the noise from stray light, and the noise from the dark. HiFLEx uses the noise measured between the traces of the echelle orders to estimate this uncertainty. The value for the uncertainties for the extracted data is calculated using
$$\sigma_{flux} = \sqrt{ flux + n_{\mathrm{pixel}} \sigma_{\mathrm{background}}^2 }\hspace{2cm},$$
with the extraction width of the order denoted by $n_{\mathrm{pixel}}$ and $\sigma_{\mathrm{background}}$ being the standard deviation of the signal in the image with the traces masked out.

In the next step a blaze corrected spectrum is derived by dividing the extracted spectrum with the blaze function (see Section~\ref{Section:blaze_function}). The blaze function is normalised to a median flux of one before the division. The uncertainties for the blaze corrected spectrum are calculated from the variation of the residuals when fitting a high order polynomial against the windowed spectrum.

The continuum of the blaze-corrected spectrum is normalised by fitting a low order polynomial to the spectrum of each order. Absorption and emission lines are masked beforehand. The noise in the spectra is measured along the continuum parts of the spectrum by fitting a low order polynomial along the measured noise. To remove the signal introduced by cosmic rays an optional sigma-clipping can be performed using a 4\,$\sigma$ threshold, while excluding the areas of absorption lines for the calculation of the scatter. 

\subsection{Storing the data}
To provide compatibility with other software packages, the reduced data is automatically stored in a variety of different ways. The pipeline stores data in the same way as the CERES pipeline \citep{2017PASP..129c4002B}, in the format provided by ESO DRS\footnote{\url{https://www.eso.org/sci/facilities/lasilla/instruments/harps/doc/DRS.pdf}}, as single files with IRAF compatible headers, and in a csv-file for the use with TERRA. The manual provides further information. If coordinates are provided by the user, the pipeline calculates barycentric correction for time and velocity \citep{2018RNAAS...2....4K} and stores the information together with other results and statistics in the image header.

\section{Results}
We have tested the package with a variety of different sets of data. This includes data taken with different detectors and different setups of a prototype spectrograph \citep{Jones20}. The package was able to trace the orders in a variety of challenging setups, which also included orders of high curvature and partly defocused orders.
Furthermore, data taken with established echelle spectrographs has been analysed, e.g. HARPS (High Accuracy Radial velocity Planet Searcher) and MRES (Medium Resolution Echelle Spectrograph at the Thai National Telescope, \citet{2018SPIE10714E..07B}). MRES does not have a data reduction package whereas HARPS data can be analysed with different packages. The resolution of the respective spectrographs is 13,000 and 115,000. 

\subsection*{Reduction of MRES spectra}

\begin{figure}
  \centering
  \includegraphics[width=0.8\textwidth]{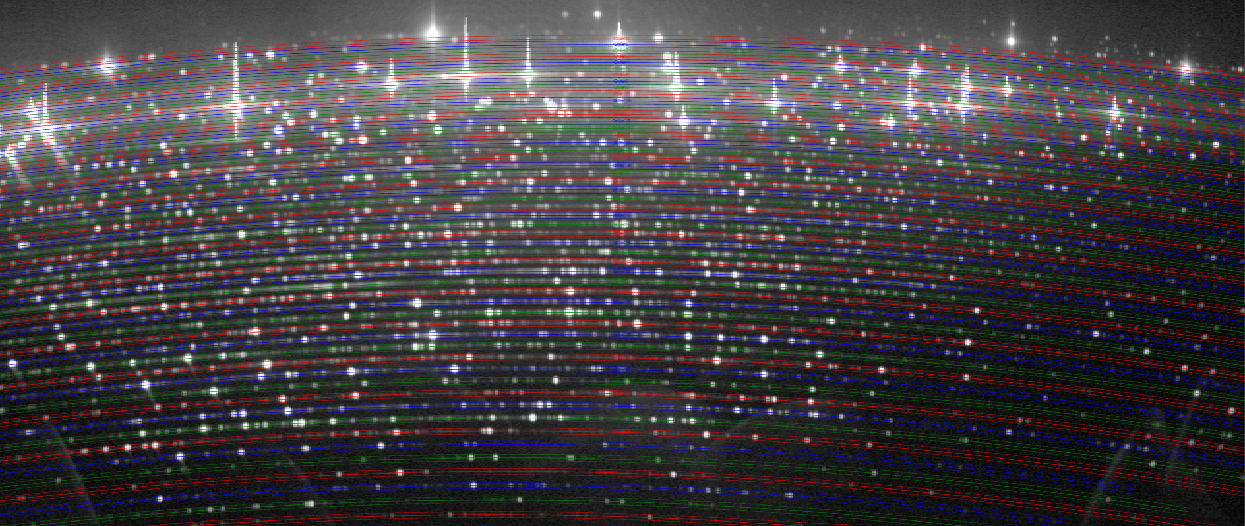} 
   \caption{Images of a long-exposed Th-Ar spectrum taken with MRES. The brightness is scaled to logarithmic. The traces of the orders are shown (alternating red, blue, green).} 
  \label{fig:screenshot_MRES_identified_traces_arc}
\end{figure}
We observed stars which might have had or in the future will have close encounters with the Solar System \citep{Tanvir18, Tanvir20} with MRES. Possible close encounters of the solar system have been found by modelling the orbits of 14,374 stars using the Gaia DR2 data. MRES spectra provide a better radial velocity determination than the Gaia DR2 results and hence allow more precise orbits. We observed some of the most likely targets from that sample between November 2018 and April 2019 (programme ID C06\_019). The observation included taking all necessary calibration data (Tungsten spectra, Th-Ar spectra with different exposure times, biases). MRES is not a stabilised instrument, hence a few times during the night Th-Ar spectra were taken to monitor the drift of the spectrograph. The detector was cooled to -100$^{\circ}$C. The exposure time for the science spectra were between 45 and 60\,minutes, with each target being observed once.

The data was reduced using the HiFLEx pipeline with the parameters described in the manual. Only bias correction was applied to the raw images. 
Figure~\ref{fig:screenshot_MRES_identified_traces_arc} shows the Th-Ar spectrum of MRES. The marked traces were identified in the master image of the Tungsten spectra. The red echelle orders have a small separation between each other due to limited diffraction by the cross-dispersing prism. However, even the slightly overlapping orders could be identified by HiFLEx.

\begin{figure}
  \centering
  \includegraphics[width=0.8\textwidth]{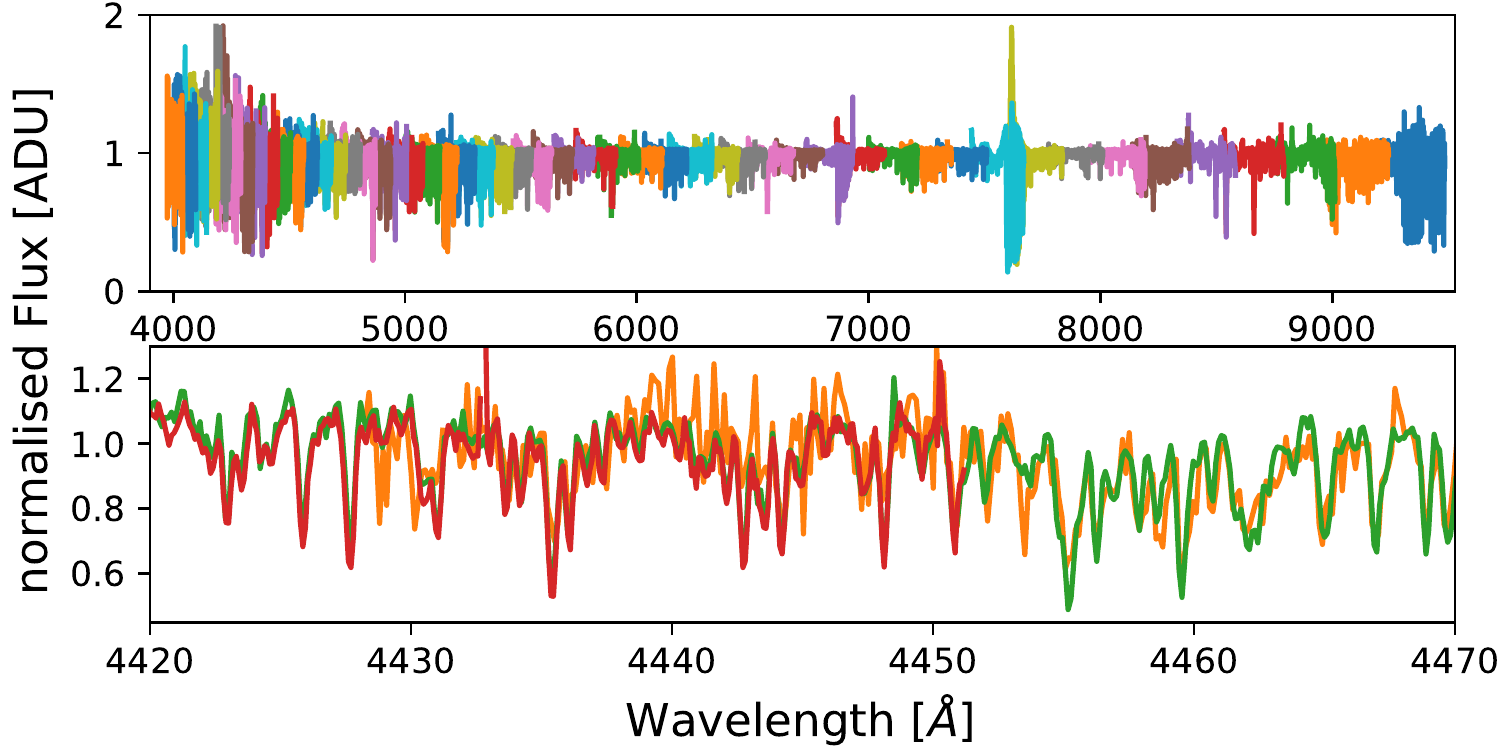}  
   \caption{Spectrum of Gaia DR2 510911618569239040, one of the target stars, for MRES after blaze correction and normalisation of the spectrum to a continuum flux of 1\,ADU. The top panel shows the light from all 52~MRES orders, with each order being in one colour. The bottom panel shows the overlap between several orders around 4450\,\AA. The spectral region plotted is fully covered by the order in green, whereas a bluer order plotted in red finishes around 4450\,\AA\ and a redder order in orange starts around 4430\,\AA. } 
  \label{fig:MRES_spectrum}
\end{figure}

After the preparation for the science extraction have been finished, the spectra of the target stars were extracted. Figure~\ref{fig:MRES_spectrum} shows the spectrum of one of our science targets (first entry in Table~\ref{Tab:MRES_RV}). The figure shows wavelength ranges with larger amounts of deep absorption lines, for example the telluric lines around 7650\,\AA. In a close-up of the overlapping orders (bottom panel in Figure~\ref{fig:MRES_spectrum}), lower flux in the blue side of the orange order and therefore lower signal-to-noise ratio is visible. That is the spectra from blue edge of an order (and the red spectrum from red side of another order) have lower signal-to-noise compared to the green spectra because grating efficiency falls away significantly away from centre of echelle orders. However, the normalisation is not affected by this.
The final radial velocities were measured only with the radial velocity code of the CERES package, as this code provides absolute RV against an intrinsic template spectrum. Table~\ref{Tab:MRES_RV} gives the radial velocity values determined in this way. Our results are consistent with the radial velocities from other sources for most stars and only in a few cases, e.g., for the targets 5295..., 2946... and 3376... do the MRES velocities indicate significantly different velocities to indicate substantially different orbits making it appropriate to obtain further follow-up observations at different epochs to further investigate the orbits of these stars. 
\begin{table}
  \caption{Radial velocity (RV) measurements based on Gaia DR2 \citep{2018A&A...616A...1G}, HiFLEx, LAMOST-DR4$^{\mathrm{a}}$ and RAVE5 \citep{2017AJ....153...75K} for the most probable close encounters of the Solar System based on the range of probable velocities (5-95\%) confidence interval, $v_{\mathrm{range}}$) for a nominal closest approach distance ($d_{\mathrm{nom}}$) at a nominal time ($t_{\mathrm{nom}}$) where negatives values are past encounters. The objects Gaia DR2 designations are given in the first column, radial velocity uncertainties are given in brackets.
  $^{\mathrm{a}}$: \url{http://dr4.lamost.org}}
  \label{Tab:MRES_RV}
 \begin{tabular}{l r c c c c c c}
  Gaia DR2     & $G$   & $d_{\mathrm{nom}}$ & $t_{\mathrm{nom}}$ & $v_{\mathrm{range}}$ & Gaia DR2 & HiFLEx-   & other \\
  designation  & [mag] & [pc]          & [Myr]         & [km/s]          & RV [km/s] & RV [km/s] & RV [km/s] \\
\hline
510911618569239040  & 8.9  &  0.43 & -2.86 & 25.9-27.1 & 26.45 (0.35)  &  26.93 (0.28) &  \\
436664033586936704  & 11.1 &  0.36 & -2.13 &-28.0-29.2 & -28.59 (0.35) & -29.33 (0.29) &   -28.94 (0.63)$^{\textrm{b}}$ \\
2906805008048914944 & 10.7 &  0.42 & -1.71 & 43.6-51.7 & 47.55 (2.37)  &  45.95 (0.29) &  48.45 (1.53)$^{\textrm{b}}$\\
5571232118090082816 & 11.8 &  0.18 & -1.17 & 81.6-83.1 & 82.18 (0.47)  &  82.08 (0.30) &  \\
5700273723303646464 & 12.0 &  0.18 & -1.64 & 36.5-39.6 & 38.05 (0.91)  &  37.31 (0.29) &  \\
5599691155509093120 & 12.2 &  0.33 & -0.84 & 71.3-78.5 & 75.01 (2.19)  &  72.15 (0.46) &  \\
52952724810126208   & 12.2 &  0.26 & -0.54 & 32.8-43.5 & 37.77 (3.35)  &  30.55 (0.79) & 25.70 (6.05)$^{\textrm{c}}$ \\
2946037094755244800 & 12.3 &  0.24 & -0.91 & 36.9-47.4 & 42.11 (3.20)  &  29.21 (0.44) &  \\
955098506408767360  & 12.4 &  0.09 & -0.74 & 34.9-42.0 & 38.52 (2.12)  &  39.04 (0.77) & 33.19 (6.31)$^{\textrm{c}}$ \\
4472507190884080000 & 12.9 &  0.36 & 1.81  &-26.7-77.7 &-52.18 (15.23) & -57.56 (0.58) &  \\
3376241909848155520 & 12.5 &  0.48 & -0.45 & 70.3-89.2 & 79.91 (5.63)  &  67.27 (0.63) &  \\

 \end{tabular}
 $^{\textrm{b}}$: RAVE5, $^{\textrm{c}}$: LAMOST-DR4
\end{table}

\subsection*{Comparison of HiFLEx extraction to other packages}
In order to more systematically benchmark the package, data analysis has been performed on publicly available data from HARPS. Here we focus on spectra taken of the asteroid Ceres and the corresponding calibration to compare results between other data reduction and radial velocity analysis packages. The data of Ceres provides a reflected solar spectrum. The data were taken in one night (2015-07-30) with HARPS at the 3.6\,m telescope at ESO's La Silla observatory. 

\begin{figure}
  \centering
  \includegraphics[width=0.8\textwidth]{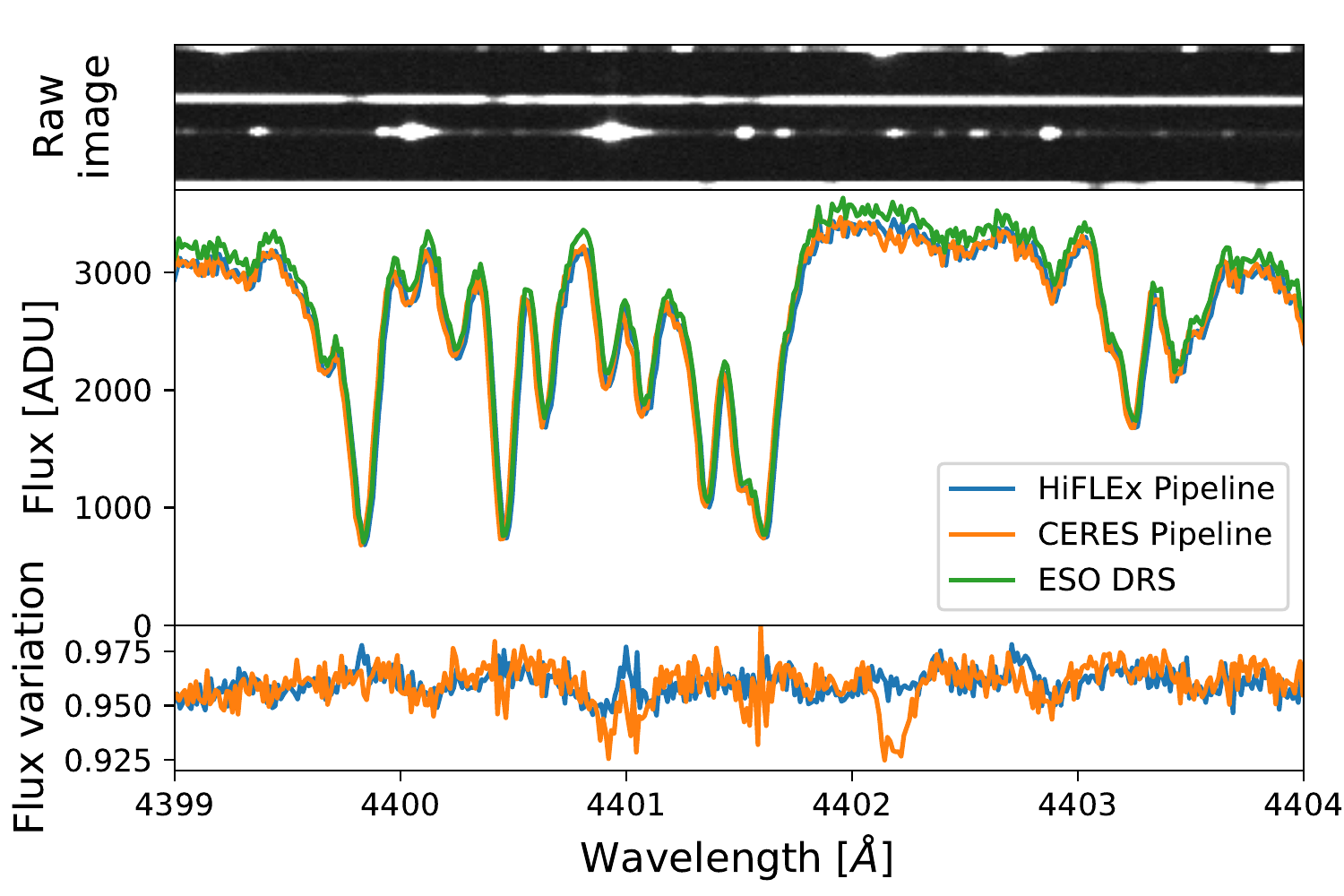}
   \caption{Example of the extracted data for the asteroid Ceres taken with HARPS. The top panel shows the raw data of order 23 of the blue chip for both the science and calibration fiber and the next closest traces. The middle panel shows the extracted spectrum of the science fiber using different packages. The ESO DRS results have been converted from e$^-$ to ADU. The CERES pipeline provides the wavelength in vacuum reference format and with barycentric correction, these two corrections were undone in order to compare with the ESO DRS wavelengths. The bottom panel shows the difference between the HiFLEx and CERES extracted spectrum, when these spectra are divided by the ESO DRS spectrum.} 
  \label{fig:spectra_comparison}
\end{figure}

For the different extraction packages tested, an area of the extracted spectrum is shown in Figure~\ref{fig:spectra_comparison} for the chosen regime in the centre of a central order on the blue chip. The variation of the normalised spectrum relative to the ESO DRS is 0.6\% for HiFLEx and 1.0\% for CERES. This variation appears to have its origin in the width of the traces. In the HiFLEx configuration file
the area to be extracted was set to where the light in the image of the flat field lamp spectrum is above 5\%$^{\ref{Footnote:url_github_manual}}$. This means that some light (with the lowest signal-to-noise ratio) outside of the area is not extracted. However, other data reduction packages use other thresholds leading to small variation in the flux levels. 
Iterative change of the parameters could be done to optimise and achieve a spectrum closer to the fixed ESO DRS pipeline. For example the curvature of the orders could be defined with a higher order polynomial or the extraction width could be changed. The largest difference between the packages appears to occur in the vicinity of bright emission lines (see top panel of Figure~\ref{fig:spectra_comparison}). HiFLEx does not correct for small scaled stray light, while the other two packages correct for this in different ways. This may account for the apparent small features at 4401\,\AA\ and 4402.2\,\AA\ in the CERES spectrum where it can be seen that the flux is few percent lower. 

\begin{figure}
  \centering
  \includegraphics[width=0.8\textwidth]{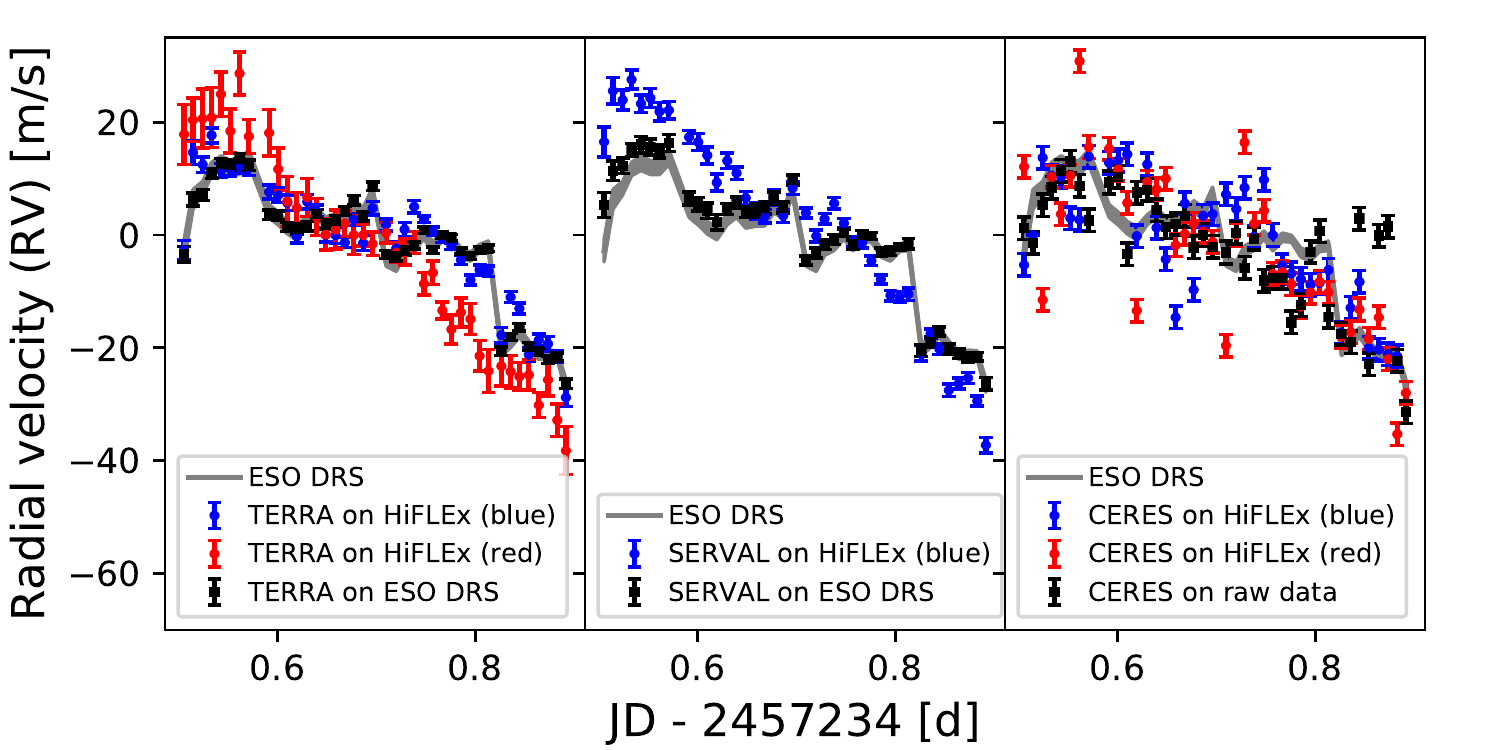}  
  \caption{Relative radial velocities of HARPS spectra of the asteroid Ceres using different packages. Each panel in turn shows the results from one package: TERRA, SERVAL, CERES. In grey the results provided by the ESO DRS are given for comparison. HiFLEx was applied to each of the two HARPS CCD individually, blue and red referring to the wavelength range of the CCD (circles). TERRA and SERVAL were applied to the extracted spectra (e2ds-files) from ESO DRS (black squares). CERES was used to reduce the raw-data and measure the RVs in this spectra (black squares). SERVAL results are not shown for red CCD.}
  \label{fig:RV_comparison}
\end{figure}

The radial velocity of the data was measured with the packages TERRA, SERVAL, and CERES. TERRA and SERVAL build a master spectrum from the data and measure the radial velocities against this spectrum. However, the barycentric velocity varied about 1\,km/s in the observation window. This means that unmasked telluric lines might not be averaged out and hence might affect the the templates built by TERRA and SERVAL. This effects the red HARPS chip in particular. On the other hand, the CERES pipeline measures the radial velocity using template spectra. It re-scales the error bars using parameters determined for each instrument by computing the RV of a high S/N spectrum when different noise levels are added \citep{2017PASP..129c4002B}. In this work, we have used the same parameters as provided by the CERES pipeline for both HARPS chips together. However, as our measurements were based on the individual chips and therefore less orders were available to determine the radial velocities, the error bars are underestimated. Furthermore, this leads to similar error bars for the blue and red chip.

Figure~\ref{fig:RV_comparison} shows the radial velocity (RV) measurements analysing the data of asteroid Ceres with different packages. In the HiFLEx extraction, the data from the two CCD chips in HARPS were analysed individually. The RV precision is lower and scatter higher, compared to packages analysing all orders from both chips at once. This is particularly the case for the red chip as the number of orders and the number of stellar absorption lines per order is lower and can be seen in the much larger error bars in the TERRA RVs. For SERVAL we excluded the RVs from the red chip completely. The RV analysis using CERES shows a higher scatter compared to the results from the other two packages. This is seen when reducing the data with CERES (black squares) and as well when determining the RV of the extracted HiFLEx spectra for each chip individually (blue and red circles).

In this comparison the extraction parameters were not optimised to the HARPS spectrograph. The flexibility of the HiFLEx package has an impact on the precision of its wavelength solution. This apparently arises because our automatic process requires that only Th-Ar lines above a $2\sigma$-threshold are used and lines that deviate from the solution are clipped. 

We want to emphasise that this comparison is very limited in scope. Figure~\ref{fig:RV_comparison} serves to illustrate that HiFLEx can obtain RVs within $\pm3$\,m/s (based on the blue chip) of sophisticated bespoke packages (ESO DRS) without any particular fine tuning of extraction parameters when using TERRA, 7\,m/s when using SERVAL, and 6\,m/s when using CERES.

\section{Conclusion}
We present the data reduction package HiFLEx that allows the extraction and analysis of cross-dispersed spectra. It provides automatic data reduction for spectrographs without a bespoke package. Preparing the reduction for a new instrument can be achieved within a working day, significantly less if the pixel to wavelength relation is provided. Nevertheless, the high flexibility of the package comes with the price of less precision, usually dominated by line identification and the corresponding wavelength solution. Testing the package on HARPS data from a single night indicates radial velocities with a scatter of 3\,m/s can be obtained without any fine tuning.

Future versions will include more sophisticated procedures for wavelength calibration and cosmic ray removal to give improved functionality and resilience, e.g. the \texttt{deepCR} package \citep{2019JOSS....4.1651Z} and \texttt{xwavecal} \cite{2019arXiv191008079B}. We also plan to expand the compatibility to further RV packages, some given in the Introduction. Further optimisation in computational efficiency are possible, e.g. in the extraction process and in parallelisation. Due to the extra steps required for the flexibility, extraction needs up to 50\% more computation time than the other tested packages for this work.

\section*{Data access}
MRES data used in this publication can be obtained from \url{https://uhra.herts.ac.uk/} under a Creative Commons Attribution license. HARPS data is available from the ESO achive.

\section*{Acknowledgments}
This work has made use of data from the European Space Agency (ESA) mission
{\it Gaia} (\url{https://www.cosmos.esa.int/gaia}), processed by the {\it Gaia}
Data Processing and Analysis Consortium (DPAC,
\url{https://www.cosmos.esa.int/web/gaia/dpac/consortium}). Funding for the DPAC
has been provided by national institutions, in particular the institutions
participating in the {\it Gaia} Multilateral Agreement.\\
RE, HRAJ and WM acknowledge support from the UK’s Science and Technology Facilities Council (STFC) [ST/R006598/1]. We thank Matthias Zechmeister for helpful discussion. We would also like to thank the anonoymous referee for their many insightful suggestions.

\bibliographystyle{plainnat}  
\bibliography{references}  

\end{document}